\documentstyle[preprint,aps]{revtex}
\begin{document}
\draft
%


\title {\Large \bf
Feshbach Resonance and Hybrid Atomic/Molecular BEC-Systems}
\author{Paolo Tommasini$^{1}$, 
Eddy Timmermans$^{1}$, Mahir Hussein$^{2,1}$
and Arthur Kerman$^{3}$\\
$^{1}$Institute for Atomic and Molecular Physics \\
Harvard-Smithsonian Center for Astrophysics 
60 Garden Street, Cambridge, MA 02138 \\
$^2$ Instituto de F\'{i}sica, Universidade de S\~{a}o Paulo, C.P. 
66318, \\
CEP 05315-970 S\~{a}o Paulo, Brazil \\
$^{3}$ Center for Theoretical Physics, Laboratory for Nuclear Science and \\
Department of Physics, Massachusetss Institute of Technology\\
Cambridge, MA 02139}
\date{\today}
\maketitle
\begin{abstract}

	The interactions responsible for Feshbach resonances in binary
atom collisions produce a condensate of molecules in the atomic BEC-system.  
We discuss the `ground state' of the condensate system and illustrate
that its properties differ quantitatively and qualitatively from the
properties of a single condensate with atom-atom interactions described by
the effective scattering length that characterizes the binary 
atomic collisions.  We show, for
example, that the tunneling of atom pairs between the atomic and 
molecular condensates
lowers the energy and can bind the condensate system, giving it
the liquid-like property of a self-determined density.  
\end{abstract}



The recently observed atomic Bose-Einstein condensates \cite{first} 
are dilute superfluid quantum gas systems and their many-body 
properties are highly interesting.
In contrast to the situation in the traditional (Helium) superfluids,
the strength of the inter-particle interactions in the atomic
condensates can vary over a wide range of values.  In particular,
the scattering length that characterizes the atomic interactions
can be negative, corresponding to an effective inter-atomic attraction.
A condensate of such bosons collapses in
the absence of an external potential, but a trapping potential
may hold a small negative scattering length condensate
in a metastable state \cite{fourth}
that can tunnel \cite{Leggett}.
Moreover, the remark that the scattering length
may be varied continuously for binary collisions
seemed to promise a `tunable' interaction
strength.  The proposed
schemes \cite{Kagan} to change the scattering length involve a resonant
scattering process with an intermediate bound molecular state
of the interacting atoms.  In the scheme that involves a 
Feshbach resonance \cite{Verhaar1}, recently achieved experimentally 
in an atomic BEC-system \cite{MITFesh},
the molecular intermediate state energy is tuned
near-resonance (i.e. near the continuum of the incident atom channel)
by varying an external magnetic field.  

	In this letter, we point out that the interactions
responsible for the Feshbach resonance in the binary atom collisions
produce an additional condensate of molecules.  
Near-resonance, the properties of the hybrid atomic/molecular condensate
differs qualitatively from the properties of a single condensate.  As a
particularly striking illustration, we will show that the tunneling of
atomic pairs between the atomic and molecular condensates can produce
a bound many-body state with the liquid-like property of a self-determined
density.

	The binary atom Feshbach resonances studied 
by varying a strong external magnetic
field in an alkali-atom trap
are hyperfine-induced spin-flip processes that bring the colliding atoms to a
bound molecular state of different electron spin. 
The interaction of two alkali atoms, their spins aligned by the magnetic field, 
is, to lowest order, described by the triplet molecular potential.  
If the magnetic field
is tuned near-resonance, the 
hyperfine interaction of
the electronic (${\bf s}$) and nuclear (${\bf i}$) spins, $V_{hf} \sim
\; {\bf s} \cdot {\bf i}$, can simultaneously flip the electronic and
nuclear spins of one of the colliding atoms, bringing the
molecular system to a bound vibrational eigenstate of a
singlet potential of nearly the same energy as the continuum
of the incident atom channel.  For the ultra-cold condensate
systems, the matrix element for the transition of
the initial state of atoms 1 and 2 in spin state $|S_{{\rm in}} \rangle$
to the bound molecular state of spin $|S_{{\rm m}}\rangle$,
can be taken to be independent of the momenta of the atoms.  
This matrix element, $\alpha$, is equal to
$\alpha = \langle S_{{\rm in}} | V_{hf}(1) + V_{hf}(2) 
| S_{{\rm m}} \rangle 
\times \int d^{3} x \;
\psi_{m}^{\ast}({\bf x}) \psi({\bf x})$, where ${\bf x}$ denotes the relative
position of atoms 1 and 2  and $\psi_{m}$ and  $\psi$
represent the bound state and continuum
wave functions of the initial and intermediate molecular states.
The continuum wave function is normalized by requiring $\psi$ to go over
into the partial s-wave component of a plane wave at large separation.
The hyperfine-induced spin-flips are then described by the following
Feshbach resonance
contribution to the Hamiltonian:
\begin{equation}
H_{FR}  = 
\alpha \int d^{3} r \;  \hat{\psi}_{m}^{\dagger} ({\bf r})
\hat{\psi}_{a} ({\bf r})
\hat{\psi}_{a} ({\bf r}) \; 
+ h.c. ,
\label{e:hp2}
\end{equation}
where ($\hat{\psi}$) and 
($\hat{\psi}^{\dagger})$ represent the usual annihilation and
creation field operators of the atoms ($\hat{\psi}_{a}, 
\hat{\psi}_{a}^{\dagger})$ and the molecules in the bound vibrational 
state ($\hat{\psi}_{m}, 
\hat{\psi}_{m}^{\dagger})$.  

	For the binary atom system, $H_{FR}$
gives a resonant contribution to the atom-atom interaction strength
$\lambda_{a}$ (where $\lambda_{a}$ is proportional to the scattering
length $a_{s}$,
$\lambda_{a} = 4 \pi \hbar^{2} a_{s} /m$): $\lambda_{{\rm eff}} =
\lambda_{a} - 2 \alpha^{2}/\epsilon$, where $\epsilon$ is the energy of the
intermediate molecular state relative to the continuum of
the incident atoms.  The value of $\epsilon$, which we shall refer to as the `detuning', linearly depends on the magnetic field $H$.  To see that,
we note that the energy difference $\Delta$ of the singlet and triplet 
continuum levels, which is
the energy gain experienced by a single atom in the spin flip process,
is proportional to $H$: $\Delta = \mu' H$.  The detuning is the difference
of the $\Delta$ and the the binding energy $E_{b}$ of the resonant singlet
molecule, so that $\epsilon = \mu' (H - H_{0})$, where $H_{0} = 
E_{b}/\mu'$ is the resonant magnetic field.
The expression for $\lambda_{{\rm eff}}$ is most easily
obtained by determining the energy shift of two atoms
with the same momentum, confined to a volume $\Omega$.
In first-order perturbation, the atom-atom interaction term,
$[\lambda_{a}/2] \hat{\psi}_{a}^{\dagger} \hat{\psi}_{a}^{\dagger}
\hat{\psi}_{a} \hat{\psi}_{a}$, contributes $\lambda_{a}/\Omega$ to
the energy shift $\Delta E$ .  In second-order perturbation theory,
$H_{FR}$ couples the initial two-atom state $|{\rm ini} \rangle$ of
energy $E_{{\rm ini}}$ to the intermediate molecular state
$|{\rm int} \rangle$ of energy $E_{{\rm int}}$, contributing
$|\langle {\rm ini} | H_{FR} | {\rm int} \rangle|^{2} / [E_{{\rm ini}}-
E_{{\rm int}}]$ $ = [-2 \alpha^{2}/\epsilon]/\Omega$ .  
The overall
shift is $\Delta E = \lambda_{{\rm eff}}/\Omega$,
where $\lambda_{{\rm eff}} = \lambda_{a} - 2 \alpha^{2}/\epsilon$ \cite{rig}. 

	In the binary collision, the intermediate state is virtual
since it does not satisfy energy conservation.  In the many-body system,
two particle energy conservation considerations do not apply
and we argue that, under the appropriate conditions,
$H_{FR}$ produces a condensate of molecules.
If all atoms are in 
the same center-of-mass state, then the molecules created in the Feshbach
resonance also occupy a single center-of-mass state, the characteristic of 
a condensate \cite{Heis}.
We can describe the dynamics of the 
atomic and molecular condensate fields, $\phi_{a}$ and $\phi_{m}$,
with an effective Hamiltonian-density \cite{ann}:
\begin{eqnarray}
&& \phi_{a}^{\ast} \left[
-\frac{\hbar^{2} \nabla^{2}}{2m} + \frac{\lambda_{a}}{2}
|\phi_{a}|^{2} \right] \phi_{a}
+ \phi_{m}^{\ast} \left[ \frac{-\hbar^{2} \nabla^{2}}{4m} + \epsilon +
\frac{\lambda_{m}}{2} |\phi_{m}|^{2} \right] \phi_{m}
\nonumber \\
&& \; \; \; \; \; \; \;  
+ \lambda |\phi_{a}|^{2} |\phi_{m}|^{2} + 
\alpha \left[ \phi_{a}^{2} \phi_{m}^{\ast} + \phi_{a}^{\ast \; 2} \phi_{m} 
\right]  \; \; \;,
\label{e:heff}
\end{eqnarray}
where the ${\bf r}$-dependence of the $\phi$-fields is understood, 
and where
$\lambda_{a}$,
$\lambda_{m}$ and $\lambda$  
denote the strengths of the atom-atom, molecule-molecule, and
atom-molecule interactions.
Having produced the molecular condensate, $H_{FR}$ describes
tunneling of atom pairs between the atomic and molecular condensates.
In the tunneling, the total number of atomic particles is conserved
if we count each molecule as two atomic particles.  We find it useful
to introduce the atomic particle density
$n({\bf r}) = n_{a} ({\bf r}) +
2 n_{m} ({\bf r})$, where $n_{a} = |\phi_{a}|^{2}$ and $n_{m} =
|\phi_{m}|^{2}$ are the atomic and molecular densities.
Note that the effective Hamiltonian depends on the phase difference
of $\phi_{a}^{2}$ and $\phi_{m}$.  In describing the static
situation, choosing $\phi_{a}$ to be real and positive, and noting
that the $H_{FR}$-contribution is minimal when $\phi_{m}$ is real, it
follows from Eq.(\ref{e:heff}) that we
can describe the statics of a homogeneous condensate system 
with the following 
energy density:
\begin{equation}
u = \frac{\lambda_{a}}{2} n_{a}^{2} + \frac{\lambda_{m}}{2} n_{m}^{2} +
\lambda n_{m} n_{a} + \epsilon n_{m} + 2 \alpha n_{a} \phi_{m} \; \; .
\label{e:edens}
\end{equation}
To minimize the energy, we introduce a variable $x$,
which is a scaled molecular field,
$\phi_{m} = x \sqrt{n/2}$.  The square of $x$ then represents the 
fraction of atoms that have been converted to molecules, 
$n_{a} = n (1-x^{2})$ and the energy per particle,
$ e = u/n$, reduces to 
\begin{equation}
e = 
\frac{\lambda_{a}}{2} n + \alpha {\sqrt{2n}} x 
+ \left\{ [-\lambda_{a} + \lambda/2 ] n + \epsilon /2
\right\} x^{2} -  \alpha \sqrt{2n} x^{3} + 
[ \lambda_{a}/2 - \lambda_{m}/4 - \lambda /2 ] n
x^{4} \; \; .
\label{e:edens2}
\end{equation}
Minimizing $e$ with respect to the molecular field parameter
$x$ within the interval $(-1,+1)$ determines the equilibrium value
of the molecular field for a given atomic particle density $n$.  
Taking $\alpha$ to be positive \cite{rep},
we reduce the problem to the minimization of
the dimensionless
function $f(x) = \left( u - \frac{\lambda_{a}}{2} n \right) / 
\alpha \sqrt{2n}$,
\begin{equation}
f(x) = x + \epsilon' x^{2} - x^{3} + \beta x^{4} \; \; .
\label{e:f}
\end{equation}
Here $\epsilon' = \left\{ \epsilon
+ n [ -2 \lambda_{a} + \lambda ] \right\}/ 2 \alpha \sqrt{2n}$,
and $\beta = n [\lambda_{a} + \lambda_{m}/4 - \lambda ] / 2 \alpha \sqrt{2n}$.

	A limit of particular interest is that of off-resonant detuning:
$\epsilon >> 2 \alpha \sqrt{2n}$, and $\epsilon$  exceeding the
single particle interaction energies,
$\epsilon >> \lambda_{a} n, \lambda_{m} n, \lambda n$. Then
$f(x) \approx
x + \epsilon' x^{2}$, with a minimum at $x=-(1/2\epsilon')$, giving
a molecular density $n_{m} \approx n^{2} [\alpha/\epsilon]^{2}$.
The ground state energy per particle
is equal to $u \approx n \lambda_{{\rm eff}}/2$, where $\lambda_{{\rm eff}}$
is the effective interaction strength
in the binary collison picture,  $\lambda_{eff} =
\lambda_{a} - 2 \alpha^{2}/\epsilon$.  
Thus the off-resonant limit ($\epsilon >> 0$) gives the same result
as the effective scattering length description.

	In contrast, 
as $\epsilon$ is lowered
near and `below' resonance 
(i.e.  $\epsilon \rightarrow 0$ or $\epsilon < 0$), the condensate system
does {\it not} exhibit resonance features.  Instead, as the detuning is 
lowered in a condensate of fixed density
\cite{meta}, the fraction of molecules increases
until at $\epsilon = - 2 \alpha \sqrt{2n}
+ n [ \lambda - \lambda_{m}/2 ]$ \cite{enperm}, 
all atoms are converted to molecules.

	The Feshbach-resonance parameter $\alpha$, proportional
to a molecular overlap matrix element, can take a wide
range of values.   Calculations
for the Feshbach resonances reported at MIT, indicate that
$\alpha \sqrt{2n}$ was of the same order of magnitude as $\lambda_{a} n$.
To be definite, we will dicuss a system of density $n_{0} \sim
10^{14} - 10^{15} cm^{-3}$ with $\alpha \sqrt{2n_{0}} = 5 \lambda_{a} n_{0}$,
roughly corresponding to the Feshbach resonance of largest width detected
in the MIT experiments \cite{MITFesh}.  If $\alpha \sqrt{2 n}$ 
significantly exceeds
the interaction energies $\lambda_{a} n, \lambda_{m} n, \lambda n$, then
$|\beta| << 1$ and the fourth order term $\beta x^{4}$ of the $f$-function
of Eq.(\ref{e:f}) may be neglected (except when the system approaches
the purely molecular condensate).  Minimizing
$f(x) \approx x + \epsilon' x^{2} - x^{3}$, we obtain a simple
analytical expression for the molecular field paramater $x$:
\begin{equation}
x = [\epsilon' - \sqrt{ \epsilon'^{2} + 3} ]/3 \; \; .
\label{e:xmin}
\end{equation}
With this result, we obtain analytical expressions for such
quantities as the chemical potential,
$\mu = (\partial u / \partial n)$, pressure, $P = n^{2} \partial e /
\partial n$, sound velocity etc... For example, in this
approximation, $\beta = n [ \lambda_{a} + \lambda_{m}/4 - \lambda]/2
\alpha \sqrt{2n} \approx 0$, we obtain
\begin{eqnarray}
P &=& \frac{\lambda_{a} n^{2}}{2}
+ n \frac{\alpha \sqrt{2 n}}{9} \left[ \epsilon' - \sqrt{
\epsilon'^{2} + 3} \right]
\left[ 1 - \frac{\epsilon'}{3} ( \epsilon' - \sqrt{
\epsilon'^{2} + 3}) \right] 
\nonumber \\
&& \; \; \; \; \; \; \; 
+ n \frac{(\lambda - 2 \lambda_{a}) n}{18}
\left[ 3 + 2 \epsilon'^{2} -2 \epsilon' \sqrt{ \epsilon'^{2} + 3} \right].
\label{e:p}
\end{eqnarray}
In Fig. (1) we show this pressure for fixed density $n = n_{0}$, and
$\lambda = 2 \lambda_{a}$ and $\lambda_{m} = 4 \lambda_{a}$
(implying $\beta = 0$, so that Eq.(\ref{e:xmin}) is exact), 
as a function of the detuning.

	The negative pressure near zero detuning 
indicates that the condensate system increases
its density.  We consider a 
condensate that is confined by fixed `walls' to a well-defined volume that is
large enough to describe the condensate by the above theory
(which is an infinite 
system treatment).  Lowering the detuning to a value of 
negative pressure, causes the system to occupy a volume less than 
that allowed by the walls. In the effective scattering
length description, which gives negative pressure when $\epsilon <
2 \alpha^{2} / \lambda_{a}$, that volume is zero: the system collapses.  
In contrast, the hybrid atomic/molecular condensate takes on a {\it finite}
volume, determined by the density
that minimizes the energy $e$.  In Fig.(2) we plot $e$ as 
a function of the relative atomic particle density $n_{s} = n/n_{0}$ for
different values of the scaled detuning $\epsilon_{s} = \epsilon /
\alpha \sqrt{2 n_{0}}$ for the same parameters as Fig.(1).  
At $\epsilon_{s} = 5$, for which the effective
scattering length description predicts zero pressure, the condensate
system still gives a positive pressure and $e$ is a monotonically
increasing function of the density.  Without the confining
walls the condensate expands indefinitely as a gas.
At $\epsilon_{s} = 4.5$, the pressure is still positive at $n=n_{0}$, but the
energy is negative at small densities, implying that the many-body system
is bound.  The binding
stems from the atom pair tunneling contribution $2 \alpha n_{a}
\phi_{m}$ to the energy density. 
At $\epsilon_{s} = 4.5$, the energy $e$ reaches a minimum
at density less than $n_{0}$. 
Thus, without walls, the condensate expands until its density reaches the
value that minimizes $e$. Since that density requires a volume
greater than that allowed by the walls, a confined condensate 
keeps its density.
At $\epsilon = 3.8$, on the other hand, the pressure is negative at $n=n_{0}$
and the density that minimizes the energy exceeds $n_{0}$: the condensate
spontaneously decreases its volume to a fraction of the
volume of confinement.  The condensate finds
its own volume and displays
the liquid-like property of a self-determined density.
Interestingly the value of this density (or, alternatively, the volume)
is determined by the detuning and can be varied by changing the magnetic
field.

	In Fig.(3), we show the 
trajectory that the condensate
follows in the plane of the detuning and density, as the detuning
is lowered adiabatically through the Feshbach resonance.
The condensate is gaseous and occupies all
of the allowed volume -- its density remains equal to $n_{0}$ until
its pressure vanishes.  Lowering the detuning further,
the condensate contracts and subsequently expands
along the curve of zero pressure.  Note that for a range of negative detuning
values, we find two densities at which the pressure vanishes.  
The lowest density at which this happens actually corresponds to
a maximim in the density, rather than a minimum, and the system could
not remain on this part of the curve.  Before the
$P = 0$ curve `turns back', the density $n$ discontinuously
changes to its  initial value, corresponding to an abrupt expansion of the
condensate to fill the maximum allowed volume.
For an atomic-trap with a smooth trapping potential, 
instead of `walls', we expect that a similar
contraction and subsequent 
expansion.

	The extent to which these very interesting phenomena can be
observed crucially depends on the relevant time scales.
Since the Feshbach resonances create molecules
in high vibrational states, the molecular lifetime, $\tau_{m}$,
is significantly shorter than the atomic lifetime.
The decay rate, $\tau^{-1}_{m}$, is proportional to the atomic 
and molecular densities,
$\tau_{m}^{-1} = c_{ma} n_{a} + c_{mm} n_{m}$, where the $c$-coefficients
are the rate coefficients due to molecule-atom ($c_{ma}$) and molecule-molecule
($c_{mm}$) collisions.  Estimates based on calculations for
hydrogen give molecular rate coefficients $\sim 10^{9} - 10^{11} cm^{3}/sec$ \cite{alex}.  Using the lower estimate, a
condensate of pure molecules of density $10^{14} cm^{-3}$ lives for
$10^{-3} sec$.  With the relevant time scale for the condensate
dynamics $\hbar/\lambda n \sim 10^{-5} {\rm seconds}$, it might be possible to
adiabatically tune $\epsilon$ near resonance, although these estimates
clearly underline the need for calculations of these quantities.
Off-resonance ($\epsilon > 0$), only a small fraction of atoms is
converted to molecules, and, although the molecules decay by colliding 
with the remaining atoms, the condensate is constantly being
replenished by tunneling from the atomic condensate.  In this limit,
$n_{m} \approx 
n_{a}^{2} |\alpha/\epsilon|^{2}$ and particle loss may 
be described by a rate coefficient that is the usual atomic rate coefficient,
$c_{aa}$, enhanced with
the rate at which atom pairs which have been converted to molecules
are being lost to molecule-atom collisions: 
$dn/dt \approx - c_{eff} n^{2}$, with $c_{eff} = c_{aa} +
2 c_{am} n |\alpha/\epsilon|^{2}$. 

	In conclusion, we have pointed out that the Feshbach resonance 
interactions can produce a condensate of molecules in an atomic BEC system.
Off-resonance, many properties of the condensate system are described
correctly in the effective scattering length description.  However,
this description does not predict the small molecular condensate, which
could be kept for times exceeding the lifetime of a pure molecular 
condensate.  Near-resonance, the tunneling between the atomic and
molecular condensates can dramatically alter the properties of the
condensate systems, a fact that we have illustrated by discussing the
appearance of a condensate phase with self-determined density.

The work of E.T. was supported by the
NSF through a grant for the
Institute for Atomic and Molecular Physics at Harvard University
and Smithsonian Astrophysical Observatory.  The work of M. H.
was supported in part by the Brazilian agencies CNPq and Fapesp.
The work of A. K. was supported in part by the U.S. Department of Energy
(D.O.E.) under cooperative research agreement DEFC02-94ER418.

\newpage
\centerline{\bf Figure Captions}
\noindent
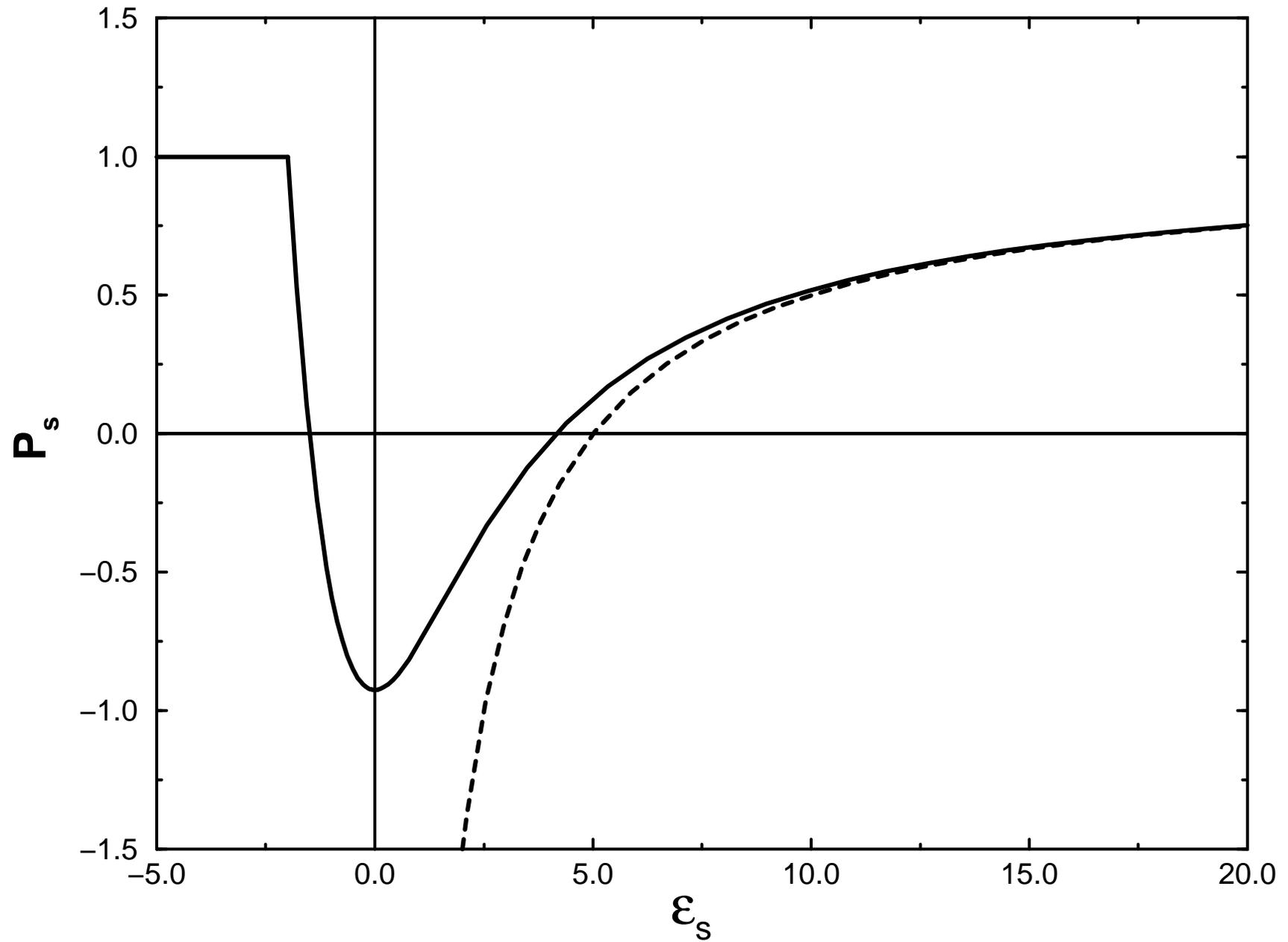
\underline{Fig.1 }:
	Scaled pressure $P_{s} = P/[\lambda_{a} n_{0}^{2}/2]$ 
as a function
of the scaled detuning $\epsilon_{s} = \epsilon /[\alpha \sqrt{2 n_{0}}]$.  
The dashed
curve shows the effective scattering length prediction.

\noindent
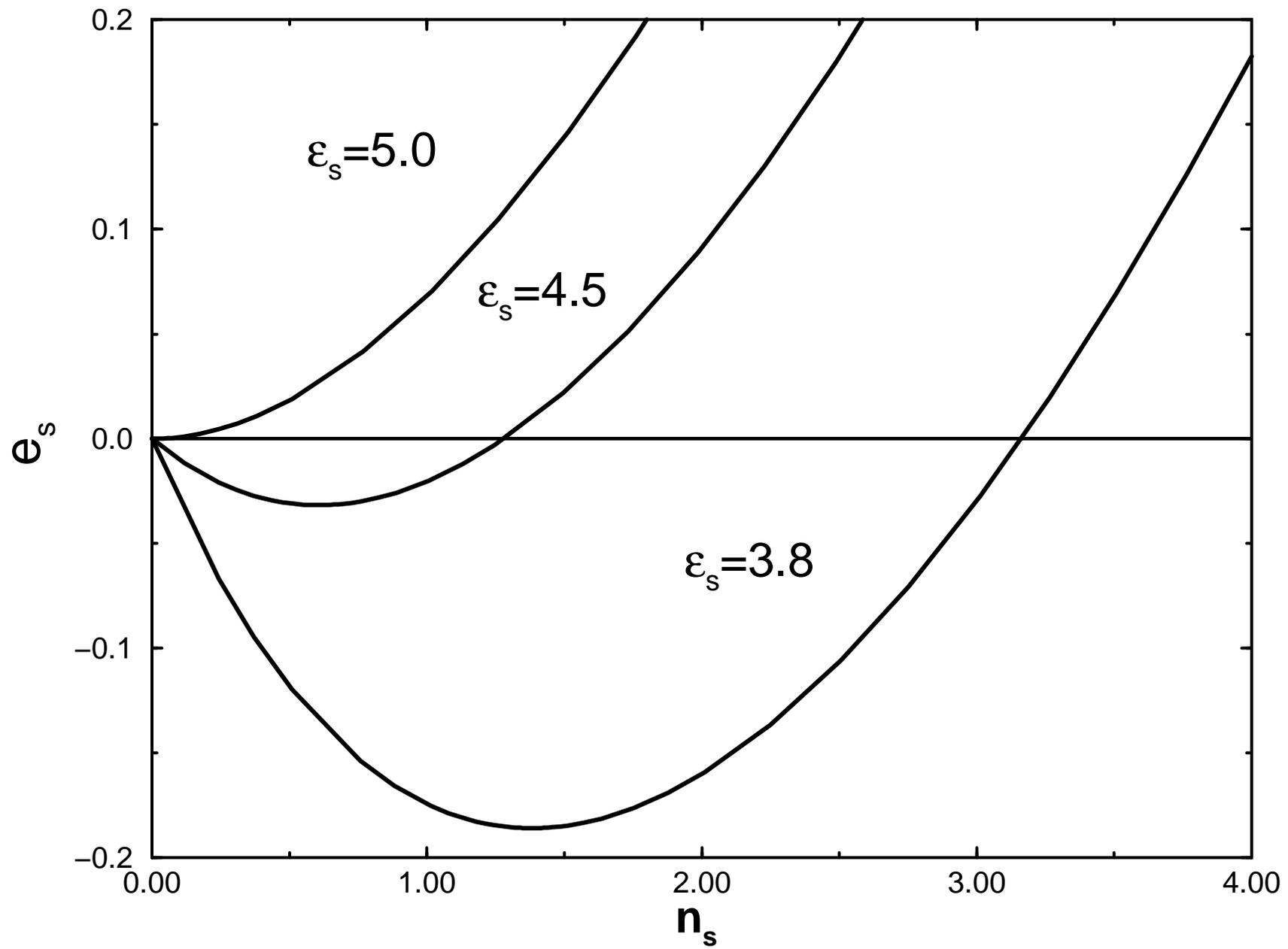
\underline{Fig.2 }:
	Plot of the energy per particle, $e_{s} = e/[\lambda_{a} n_{0}/2]$, 
as a function of the scaled density $n_{s} = n/n_{0}$ for the 
detunings shown on the figure.  Negative $e_{s}$ implies a bound
many-body state.

\noindent
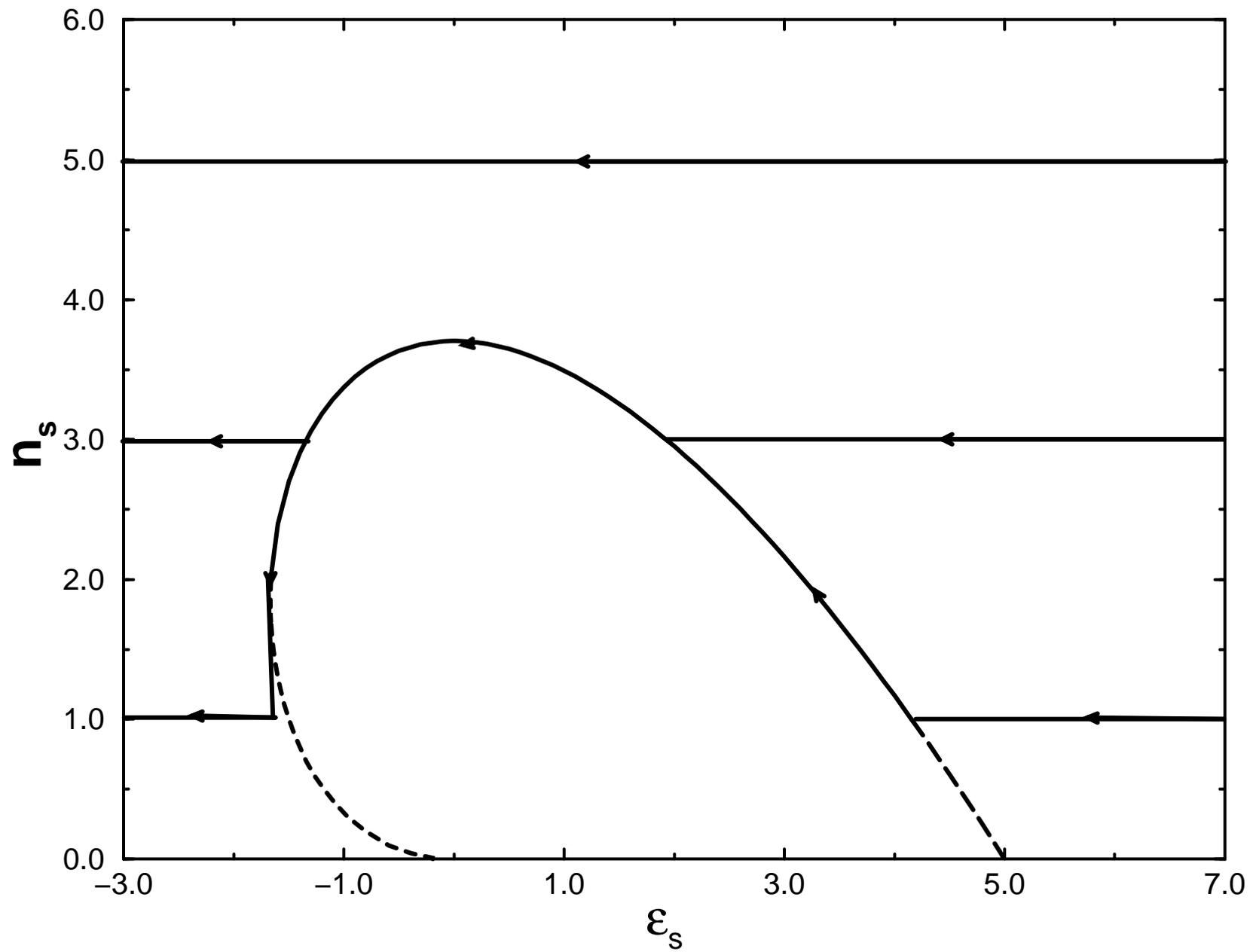
\underline{Fig.3}:
	Plot illustrating the density variation of a condensate, 
confined by fixed walls, as the detuning 
$\epsilon_{s} = \epsilon /[\alpha \sqrt{2 n_{0}}]$,
is adiabatically lowered  through the Feshbach resonance.
For the lowest initial density, the system contracts and expands 
along the curve of zero pressure, shown
in dashed line.  Before
returning to its initial density, the system undergoes
an abrupt change in the density.  Similarly, 
the system of medium initial density contracts and expands but does not
change its density abruptly.
The density of the system of highest initial density remains constant.


\begin{thebibliography}{99}
\bibitem{meta} 	We assume here that $\beta < -1$.
Otherwise a discontinuity in $x$ may occur as we shall discuss in a
more comprehensive treatment.
\bibitem{Heis} In a more
systematic approach, one may take the expectation value of
the Heisenberg equation for the field-operators.  
Denoting the fields
by $\phi = \langle \hat{\psi}\rangle$, the molecular field 
 $\phi_{m} = \langle \hat{\psi}_{m} \rangle $ is `driven' by a source term 
$\alpha \langle \hat{\psi}_{a} \hat{\psi}_{a} \rangle 
\approx \alpha \phi_{a}^{2}$ and $\phi_{m}$ does not vanish.
\bibitem{rep} Negative $\alpha$ is equivalent to 
replacing $x$ by $-x$
\bibitem{enperm} Calculating the energy gain of converting
molecules to atoms in a purely molecular condensate, we obtain
$\epsilon + 2 \alpha \sqrt{2n} - n [ \lambda - \lambda_{m}/2]$ per molecule
that is converted.  If this energy is positive, the molecular condensate
lowers its energy by producing atoms.
\end{thebibliography}
\end{document}